\documentstyle[preprint,aps,epsfig]{revtex}
\newcommand{\lsim}{\mathrel{\mathop{\kern 0pt \rlap
  {\raise.2ex\hbox{$<$}}}
  \lower.9ex\hbox{\kern-.190em $\sim$}}}
\newcommand{\gsim}{\mathrel{\mathop{\kern 0pt \rlap
  {\raise.2ex\hbox{$>$}}}
  \lower.9ex\hbox{\kern-.190em $\sim$}}}
\newcommand{\be}     {\begin{equation}}
\newcommand{\ee}     {\end{equation}}
\newcommand{\bea}     {\begin{eqnarray}}
\newcommand{\eea}     {\end{eqnarray}}
\begin{document}
\preprint{
\vbox{ \hbox{SNUTP\hspace*{.2em}01-029}
}}
\title{ 
Brane Fluctuation 
and anomalous muon magnetic moment
}

\author{
Seong Chan Park\footnote{schan@mulli.snu.ac.kr} 
~and ~~ H.~S.~ Song\footnote{hssong@physs.snu.ac.kr} 
}
\vspace{1.5cm}
\address{
Center for Theoretical Physics and
School of Physics, Seoul National University,
Seoul 151-742, Korea
}

\maketitle
\thispagestyle{empty}
\setcounter{page}{1}

\begin{abstract}
\noindent 
We study the effects of extra dimensions on the muon anomalous
magnetic moment with brane fluctuation.   
Since the coupling is naturally suppressed 
if brane fluctuation is considered by exponential softening
factor for heavier states, 
the contribution from the whole Kaluza-Klein graviton tower 
is shown to be finite.
The recent BNL E821 result is accomodated with
$D=4+\delta$ dimensional gravitational scale, $M_D$,
in the range of  $M_D \simeq 1.0 - 5.1$ TeV ($\delta=2$),
and $M_D \simeq 1.0 - 8.0 $ TeV ($\delta=6$) with the 
brane tension parameter $f=(4\pi^2 \tau )^{1/4}$,
in the range $f = 1 - 10 $ TeV.

\end{abstract}
\vskip 0.5cm
\noindent
PACS number(s): 11.30.Pb, 11.30.Er

\newpage

The report of measurement of the anomalous $g$ value for the positive muon
from Brookhaven AGS experiment 821, based on data collected in 1999,
has attracted great interests \cite{BNL}.
Combining recent theoretical and experimental uncertainties,
the new world average shows $2.6 \sigma$ deviation from 
the standard model(SM) prediction. It may be the first evidence 
that the standard model must be extended by new physics at TeV scale.
The reported result is given as
\be
\delta a_\mu \equiv a_\mu (exp) - a_\mu (SM) =43(16) \times 10^{-10},  
\ee
where $a_\mu (exp)$ is world averaged value.
Since four times larger data is already recorded in 2000 and measurements
with negative muon are undertaking now, it is very interesting situation
to consider the implications of the speculative advocates of the new
phycis on the anomaly.
Most popular approaches include weak scale supersymmetry, 
large or warped extra dimensional scenario, extension of the
gauge structure and many other alternatives (see e.g.,\cite{Nath}).
In this paper, we consider the extra dimensional scenario suggested
by Arkani-Hamed et.al. with brane fluctuation. 

If all the matter fields are confined on the brane-world, 
1-loop contribution of the Kaluza-Klein excitations of the graviton
is known to explain the recent BNL anomaly\cite{Graesser},\cite{Park},
\cite{Song}. 
(The case in which SM fields propagate in the bulk
is also studied \cite{Agashe} 
and see \cite{Contino:2001nj} for general arguments for
graviton loops and brane observables.) 
We note that in both cases-with large or
warped extra dimensions- physical cutoff at weak scale should be
imposed to get finite results from the non-renormalizable theory.
Even these truncation procedures are not un-natural in treating
effective theory \cite{Sundrum}, it seems not generally true that the exactly
same expressions for the muon anomalous moment remains valid
until the cutoff scale. In this letter, we study the another possibility
to get rid of divergence by considering brane fluctuation\cite{Bando},
\cite{Kugo:2001mf}, \cite{Wells},\cite{Park:2001hj}. 

We take the case that all the gauge and fermion fields of SM are 
confined on 3-brane and only the gravitational field can propagate
through the bulk. In that case, $D=4+\delta$ dimensional gravitational
scale, $M_D$, is related to our four dimensional Planck scale $M_P$
by $M_P^2 = M_D ^{2+\delta} R^\delta $. Where $R$ is the size of the
extra dimension. From this relation, we can understand weakness of
gravity or largeness of Planck energy scale: if the size of the extra
dimesion is large enough, the Planck scale also is very big.
It is phenomenologically interesting case, if $M_D$ is as small 
as TeV scale.

The gravitational interaction with SM fields which are confined on
the fluctuating brane can be described by the action:
\be
S \supset -\frac{1}{M_P}\int d^D x T_{\mu\nu}(x) g^{\mu\nu}(x,\vec{y})
                         \delta^D (\vec{y}-\vec{\phi}(x))
\ee
where the Nambu-Goldstone boson $\vec{\phi}(x)$, which came 
from the spontaneous
translational symmetry breaking, represent the brane fluctuating
in the $\vec{y}$ direction at point $x$ in our 3-brane.
The Nambu-Goto action with tension $\tau=f^4 / 4\pi^2$ describe the dynamics of 
$\vec{\phi}(x)$:
\be
\int d^4 x (-\tau + \frac{\tau}{2} 
\partial_\mu \vec{\phi}(x) \cdot \partial^\mu \vec{\phi}(x)
+ \cdots )
\ee
on the flat background.
After expanding the gravitational field around the compact extra 
dimension as
\be
g_{\mu\nu}(x,\vec{y})= \sum_{\vec{n}} 
g_{\mu\nu}^{\vec{n}}(x) e^{i \vec{n}\cdot \vec{y}/R},
\ee
and taking normal ordering the exponential in perturbation framework,
the interaction action includes an exponential `softening factor':
\be
S \supset -\frac{1}{M_P}\sum_{\vec{n}} \int d^4 x
e^{ -\frac{1}{2} m_{\vec{n}}^2 \Delta } 
g_{\mu\nu}^{\vec{n}}(x) T_{\mu\nu}(x) 
\label{action} \ee
where $m_{\vec{n}}^2 = \vec{n}\cdot \vec{n} / R^2 $ are the mass of the
KK modes and  $\Delta$ is the free propagator of $\vec{\phi}(x)$.
If brane tension $\tau$ is given, $\Delta$ is understood as
\be
\Delta \equiv <\phi(x)\phi(y)>|_{(x-y)^2\rightarrow M_D^{-2}}
\simeq \frac{M_D^2}{f^4} 
\ee 
since the present effective theory is valid only at scales 
lower than $M_D$.
Note that the mass dimension of $\Delta$ is $-2$ since $[\tau] = -4$.
Eq. (\ref{action}) clearly shows that if the effect of brane fluctuations
is correctly included, the coupling of the higher KK gravitons are
exponentially suppressed.

Now let us consider the effects of the KK graviton from the fluctuating
brane on $a_\mu$. The anomalous magnetic moment of the muon is the
coefficient of the operator 
$(e/4M_\mu) \bar{\mu} \sigma^{\alpha\beta} \mu F_{\alpha\beta}$, 
where $\sigma^{\alpha\beta}= (i/2) [\gamma^\alpha, \gamma^\beta]$ is
the Lorentz generator for spin-1/2 spinors.
By considering loop induced connection to $\mu\mu\gamma$ vertex, we can
calculate this term. At 1-loop order, a few Kaluza-Klein graviton mediated
diagrams can contribute to $(g-2)$ of the muon. 
Usually, radion contribution to anomalous magnetic moment is 
less than one order smaller
than that of KK graviton, we just ignore that effect in this study.

It is convenient to express the contribution of the KK gravitons
as
\be
\delta a^{KK} = \frac{1}{16\pi^2} (\frac{M_\mu}{M_P})^2 {\cal A},
\ee
where ${\cal A}$ is essentially effective degrees of freedom of 
contributing KK gravitons and it could be obtained by summing over
all relevant contributions of all the relevant KK states in the 
Feynman diagrams (see Fig.1). 
The factor $(1/16 \pi^2)$ is the usual loop factor and the suppression
factor $(M_\mu/M_P)^2$ came from the gravitational coupling strength.

In the case of rigid brane, ${\cal A}$ can be approximated as
\be
{\cal A}({\rm Rigid}) 
\approx 5 \int ^{N_\Lambda}  n^{\delta -1} \Omega_\delta d n
\ee
at the limit of our intersts:$(M_\mu /M_{KK})^2 \rightarrow 0$.
Here 5 comes from the non-decoupling contribution from the each KK graviton
mode and $\Omega_\delta= 2 \pi^{\delta/2}/\Gamma(\delta/2)$ is solid angle
in $\delta$-dimension space.
$N_\Lambda$ denotes the maximum quntum number of KK state at the
cut-off scale $\Lambda \sim M_D$ such that $N_\Lambda \sim M_D R$.
After integration, the final form of the $\delta a^{KK}$ from the
rigid brane is
\be
\delta a^{KK} ({\rm Rigid}) \approx \frac{1}{16 \pi^2} 
\frac{10 \pi^{\delta/2}}{\Gamma(\delta/2)}(\frac{M_\mu}{M_D})^2
\ee
and this result is same with the result of \cite{Graesser} 

We now consider the case that brane fluctuation is included.
With brane fluctuation, exponential softening factor naturally
suppress the effects from all the higher KK modes. The ${\cal A}$
can be casted as
\be
{\cal A}({\rm Fluctuating}) \approx 5 
\int n^{\delta-1} \Omega_\delta e^{-n^2 \Delta/ R^2 } dn.
\ee
Note that there are two vertex points for gravitational coupling
in each diagrams and so doublly suppressed factor appears in the 
above equation.
By defining dimensionless variable ${\cal L}^2 \equiv R^2 /\Delta $,
the integration gives simply gamma function as 
\bea
{\cal A}({\rm Fluctuating}) 
&&\approx 5 {\cal L}^\delta \Omega_\delta \int x^{\delta-1} e^{- x^2} dx 
\nonumber \\
&&= \frac{5}{2} {\cal L}^\delta \Omega_\delta \Gamma(\delta/2).
\eea
Note that in this case we did not introduce the cutoff at the integration
region. But we still need to 
introduce the cutoff when we consider the free propagtor
for the fluctuating field. 
Finally, the total contribution to anomalous magnetic moment of muon
from the KK graviton modes from the fluctuating brane can be approximated
as
\bea
\delta a_\mu ^{KK} ({\rm Fluctuating})
\approx \frac{5}{16\pi^2} \pi^{\delta/2}
 (\frac{f^\delta M_\mu}{M_D^{\delta+1}})^2. 
\eea
The result is sensitive to the number of extra dimensions ($\delta$).

As examples we explicitly show the expressions for the case $\delta=2$
and $\delta=6$ respectively. 
\bea
&&\delta a_\mu^{KK} \simeq 11.1 \times 10^{-10} (\frac{f}{\rm TeV})^4 
(\frac{\rm TeV}{M_D})^{6} ~~~(\delta =2) \\
&&\delta a_\mu^{KK} \simeq  11.0 \times 10^{-9} (\frac{f}{\rm TeV})^{12}
(\frac{\rm TeV}{M_D})^{14} ~~~(\delta =6).
\eea

If we ascribe the recent BNL anomaly to the KK graviton contribution,
for $f/{\rm TeV}$ in the range 1-10, then $M_D\simeq 1.0 - 5.1$ TeV if 
$D=6$ and $M_D\simeq 1.0-8.0 $ TeV if $D=10$. The Fig.2 describes
the allowed parameter space $(f, M_D)$ for the muon anomaly. The lower
side curves show the allowed region for $D=6$ and upperside curves
show the region for $D=10$. 

In summary, we have studied the effects of Kaluza-Klein tower of graviton
on the $(g-2)$ factor of muon with brane fluctuation regularization 
at 1-loop level.
After normal ordering the fluctuation 
($\sim e^{i \vec{n}\cdot \vec{\phi}(x)/R }$), we obtain the exponentially
suppressed factor ($\sim exp(-m_{KK}^2 \Delta/2)$ for heavier states. 
By this suppression, we get the finite result for muon anomalous magnetic
moment without simple neglecting procedure beyond the cutoff scale.
The recently reported deviation in anomalous magnetic moment is possibly
accomodated in the corresponding parameter space in the range of
\bea
1.0 \leq &&M_D/{\rm TeV} \leq 5.1 ~~~(D=6) \\
1.0 \leq &&M_D/{\rm TeV} \leq 8.0 ~~~(D=10)
\eea
with the tension parameter is chosen to be $f= 1-10$ TeV.

In the Randall-Sundrum model, our brane-world is set to have negative
tension \cite{RS} and such 
regularization does not work. But there is still a possibility
with brane thickness effect.

\acknowledgments
The work was supported in part by the BK21 program and in part by
the Korea Research Foundation(KRF-2000-D00077). 


%
\begin{figure}[htb]
\vskip 2.5cm  
\begin{center}
 \hbox to\textwidth{\hss\epsfig{file=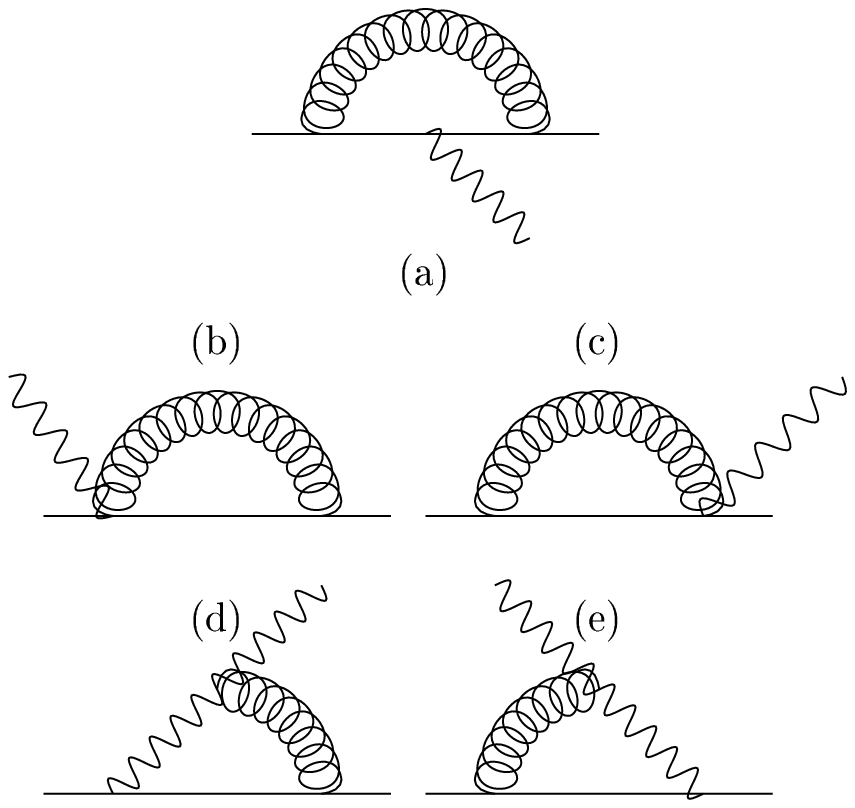,width=4cm,
                    height=4cm}\hss}
 \end{center}
 \vskip 0.5cm
\caption{\it The Kaluza-Klein mediated 1-loop diagrams for (g-2)of the muon 
. The solid and wavy line denotes muon and photon on-shell. 
The spring-shape lines denote spin-2 Kaluza-Klein states. } 
 \label{fig:fig1}
\end{figure}
\begin{figure}[htb]
\begin{center}
 \hbox to\textwidth{\hss\epsfig{file=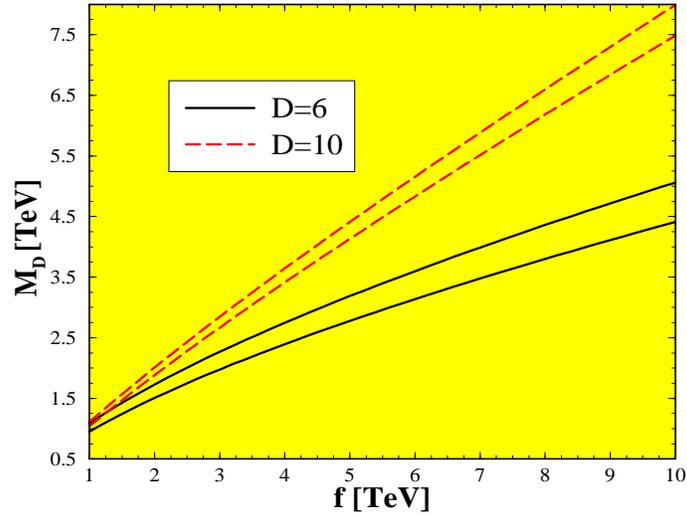,width=9cm,
                    height=7cm}\hss}
 \end{center}
 \vskip 0.5cm
\caption{\it 
The allowed parameter space ($f  , M_D$) to
compensating recent BNL E821 result of the muon anomalous
magnetic moment. The (black)solid line and the (red) dotted line
denotes the cases for $\delta=2$ (or $D=6$) and $\delta=6$(or $D=10$)
respectively.  
} 
 \label{fig:fig2}
\end{figure}

\end{document}